\documentclass{PoS}

\title{Gamma-ray emission from the black hole's vicinity in AGN}

\ShortTitle{Gamma-ray emission from the black hole's vicinity}

\author{\speaker{Frank M. Rieger}\thanks{Heisenberg Fellow}  \, and Grigorios Katsoulakos\\
        ZAH, Institute of Theoretical Astrophysics, Heidelberg University, Philosophenweg 12, 69120 Heidelberg,
        Germany, and\\ 
        Max-Planck-Institut f\"ur Kernphysik, Saupfercheckweg 1, 69177 Heidelberg, Germany\\ 
        E-mail: \email{frank.rieger@mpi-hd.mpg.de}}

\abstract{Non-thermal magnetospheric processes in the vicinity of supermassive black holes have attracted 
particular attention in recent times. Gap-type particle acceleration accompanied by curvature and Inverse 
Compton radiation could in principle lead to variable gamma-ray emission that may be detectable with current 
instruments. We shortly comment on the occurrence of magnetospheric gaps and the realisation of different
potentials. The detection of rapid variability becomes most instructive by imposing a constraint on possible 
gap sizes, thereby limiting extractable gap powers and allowing to assess the plausibility of a magnetospheric
origin. The relevance of this is discussed for the radio galaxies Cen\,A, M\,87 and IC\,310. The detection of 
magnetospheric gamma-ray emission generally allows for a sensitive probe of the near-black-hole region
and is thus of prime interest for advancing our understanding of the (astro)physics of extreme environments.}

\FullConference{7th Fermi Symposium 2017\\
		15-20 October 2017\\
		Garmisch-Partenkirchen, Germany}

\begin{document}

\section{Introduction}
The non-thermal processes occurring in the vicinity of supermassive black hole have attracted particular attention in recent 
times, e.g., \cite{lev11,rie11,bro15,pti16,hir16,lev17}. The strong electromagnetic fields around rotating black holes are often thought 
to facilitate efficient (one-shot) particle acceleration to very high energies, in the case of hadrons even up to the highest cosmic 
ray energies ($\sim 10^{20}$ eV), cf. \cite{rie11} for a review. This process is naturally accompanied by gamma-ray production 
via curvature emission (the radiation of charged particles following curved magnetic fields) and inverse Compton up-scattering 
of ambient (accretion disk) soft photons. Provided suitable conditions are present, the close black hole environment could enable 
high power extraction and account for rapid variability on timescales of $\sim r_g/c=1.4~(M_{BH}/10^9\,M_{\odot})$ hr and shorter 
\cite{alek}. A characteristic feature in this context is the occurrence of magnetospheric gaps close to the black hole.

\section{The Occurrence of Magnetospheric Gaps}\label{gap}
\begin{figure}[htbp]
\begin{center}
\vspace*{-1.5cm}
\includegraphics[height=80mm]{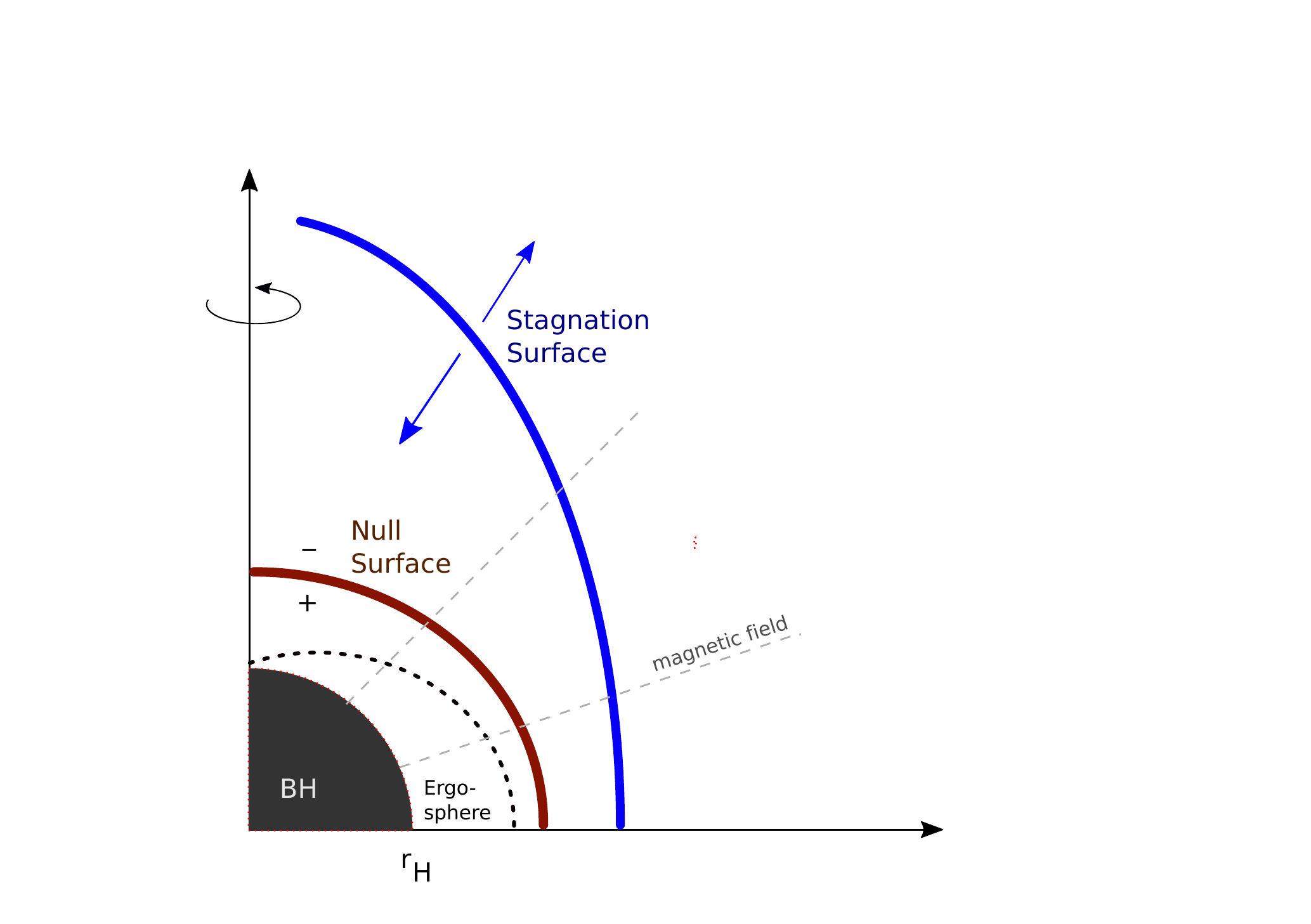}
\caption{Illustration of the possible location and structure of magnetospheric "vacuum" gaps in rotating black hole magnetospheres. 
The brown line denotes the null surface across which the charge density changes sign, and the blue line gives the stagnation surface 
from which stationary MHD flow starts.}\label{fig1}
\end{center}
\end{figure}

Unscreened parallel (magnetic-field-aligned) electric field components $E_{||}$ (so-called "vacuum gaps") could occur in at 
least two places, the {\it null surface} (NS) and the {\it stagnation surface} (SS), e.g., \cite{glo14,hir16,lev17}, see Fig.~\ref{fig1}. 
The former (NS) designates the (potentially quasi-spherical) region where the generalised Goldreich-Julian charge density 
$\rho_{GJ}$ vanishes, changing sign across it. This happens close to the location where the field line rotation frequency 
$\Omega_F$ equals the Lense-Thirring angular frequency $\omega(r)$, i.e., usually on radial scales $r \sim r_g \equiv GM/c^2$. 
In order for the black hole magnetosphere to be force-free (vanishing $E_{||}$) the real charge density $\rho_e$ should 
corresponds to $\rho_{GJ}$. As $\rho_{GJ}$ changes signs across the null surface, $\rho_e$ is required to have opposite signs 
on opposite sites, hence a parallel electric field component could easily arises around the null surface \cite{hir16}. The stagnation 
surface, on the other hand, naturally occurs in an MHD outflow driven by a rotating (Kerr) black hole and designates the surface 
that separates plasma motion inwards (inflows) due to the gravitational field from plasma motion outwards (outflows) above it. 
Plasma would need to be continuously replenished to maintain a charge density $\rho_e \geq \rho_{GJ}$ and allow for a general 
(time-averaged) force-free MHD description. The stagnation surface is in general non-spherical (with a prolate shape) and located 
inside the (outer) light cylinder, typically on radial scales of some $r_g$.

\section{The Conceptual Relevance of Magnetospheric Gaps}
It is usually believed that the magnetospheric structure of astrophysical black holes can be reasonably approximated (at 
least in a time-averaged sense) by an electromagnetic force-free solution. Such a configuration is known to facilitates an 
efficient electromagnetic extraction of the rotational energy of the black hole \cite{bla77}. For this to be possible, however, 
some quasi-steady electric currents that pervade the magnetospheres as sources of the magnetic field and that are carried 
by charged particles flowing through them, need to be sustained despite the presence of inflows and outflows. 
If we suppose, to the contrary, that under-dense ($\rho_e < \rho_{GJ}$) regions (gaps) are formed in which a parallel electric 
field is established, see Sec.~\ref{gap} above, this does not necessarily invalidate overall force-freeness. In fact, if the 
potential is sufficiently strong as expected around supermassive black holes, a single particle entering the gap can be 
accelerated to very high energy (up to Lorentz factors $\gamma\sim 10^{10}$) in the parallel field $E_{||}$, emitting curvature 
and inverse Compton gamma-ray photons on its way. In the presence of an ambient photon field, an electron-positron pair 
production ($\gamma\gamma\rightarrow e^+e^-$) cascade is triggered, that leads to vacuum breakdown and the formation 
of highly conducting pair plasma, i.e., to enough charges ($\rho_e=\rho_{GJ}$) to annihilate the parallel electric field component 
and ensure gap closure ($\vec{E}\cdot\vec{B}=0$), e.g., \cite{lev11}. As a consequence, the magnetic field lines can be 
considered as nearly orthogonal to the $\vec{E}$ field lines, with just enough $\vec{E}\cdot\vec{B}$ remaining to produce 
sparks of electron-positron pairs and keep the magnetosphere filled with plasma \cite{tho86}. This suggests that an 
electromagnetic force-free solution provides a reasonable approximation to the time-averaged structure of such a 
magnetosphere \cite{bla77}.\\ 
While the qualitative picture seems evident, the resulting non-thermal emission features (e.g., maximum particle energy, 
dominant emission mechanism, radiative window, gap power) will depend on the details of the magnetospheric set-up 
(boundary conditions), and different realisations of the electric field and potential are in principle conceivable (and actually
encountered, see e.g., \cite{kat17}) as motivated below.

\subsection{Gap Potential Realisations}
In its simplest (one-dimensional, non-relativistic) form the gap electric field along $s$ in the presence of a non-zero charge
 charge density $\rho_e$ can be determined from Gauss' law
\begin{equation}\label{Gauss}
 \frac{dE_{||}}{ds}= 4\pi (\rho_e -\rho_{GJ})\,, 
\end{equation} and the electrostatic potential from 
\begin{equation}
\frac{d\Phi_e}{ds}= -E_{||}\,,
\end{equation} so that the relevant voltage drop becomes $\Delta \mathcal{V}_{\rm gap}= \Phi_e(s=h)-\Phi_e(s=0)$, where $h$ 
denotes the characteristic gap height. Different solutions are however obtained dependent on which boundary conditions are 
considered to be realised as shown in ref.~\cite{kat17}: if  $\rho_e \ll \rho_{GJ}$ ({\it case a}), for example, then 
\begin{equation}\label{potential1}
\Delta \mathcal{V}_{\rm gap,a}=\Phi_0 \,\left(\frac{h}{r_g}\right)^2\,,
\end{equation} with $\Phi_0\simeq \Omega_F r_g^2 B_H/c$, where $B_H$ corresponds to the strength of the normal magnetic 
field component threading the horizon, while for $\rho \sim \rho_{GJ}$ ({\it case b}) one instead finds
\begin{equation}\label{potential2}
\Delta \mathcal{V}_{\rm gap,b}\simeq \frac{1}{6} \Phi_0 \,\left(\frac{h}{r_g}\right)^3\,,
\end{equation} i.e., a scaling $\propto h^3$ where the power index is increased by one compared to the previous one.

\subsection{Associated Gap Luminosities}
Given the anticipated potential strengths around supermassive black holes, a significant amount of the non-thermal emission of 
magnetospheric gaps is expected to occur in the high and very high energy (VHE) gamma-ray domain, e.g., \cite{rie11,lev11,hir16,
pti16}. The different gap potential noted above, however, result in different expectations for the maximum gap luminosity $L_{\rm gap}
\simeq n_c~V_{gap}~dE_e/dt$ (with characteristic gap volume $V_{\rm gap} \propto r_g^2 h$) for a gap of height $h$. As we show in
ref.~\cite{kat17} the maximum extractable gap power is in general proportional to the classical Blandford-Znajek jet power, $L_{\rm BZ} 
\propto r_g^2 B_H^2 \propto \dot{m} \,M_{BH}$ (with $B_H \propto \dot{m}^{1/2}$), and a sensitive function of the gap height $h$, 
\begin{equation}\label{power}
L_{\rm gap} \simeq \eta_{\beta} L_{\rm BZ}\,\left(\frac{h}{r_g}\right)^{\beta}\,,
\end{equation} where the power index $\beta\geq 1$ is dependent on the respective gap-setup, i.e., $\beta=2$ with $\eta_{2}=1$ 
for the noted {\it case a} and $\beta=4$ with $\eta_{4}=1/6$ for the {\it case b} above. 

\section{The Phenomenological Relevance of Magnetospheric Gaps}\label{ref}
\begin{figure}[b]
\begin{center}
\vspace*{-0.5cm}
\includegraphics[height=80mm]{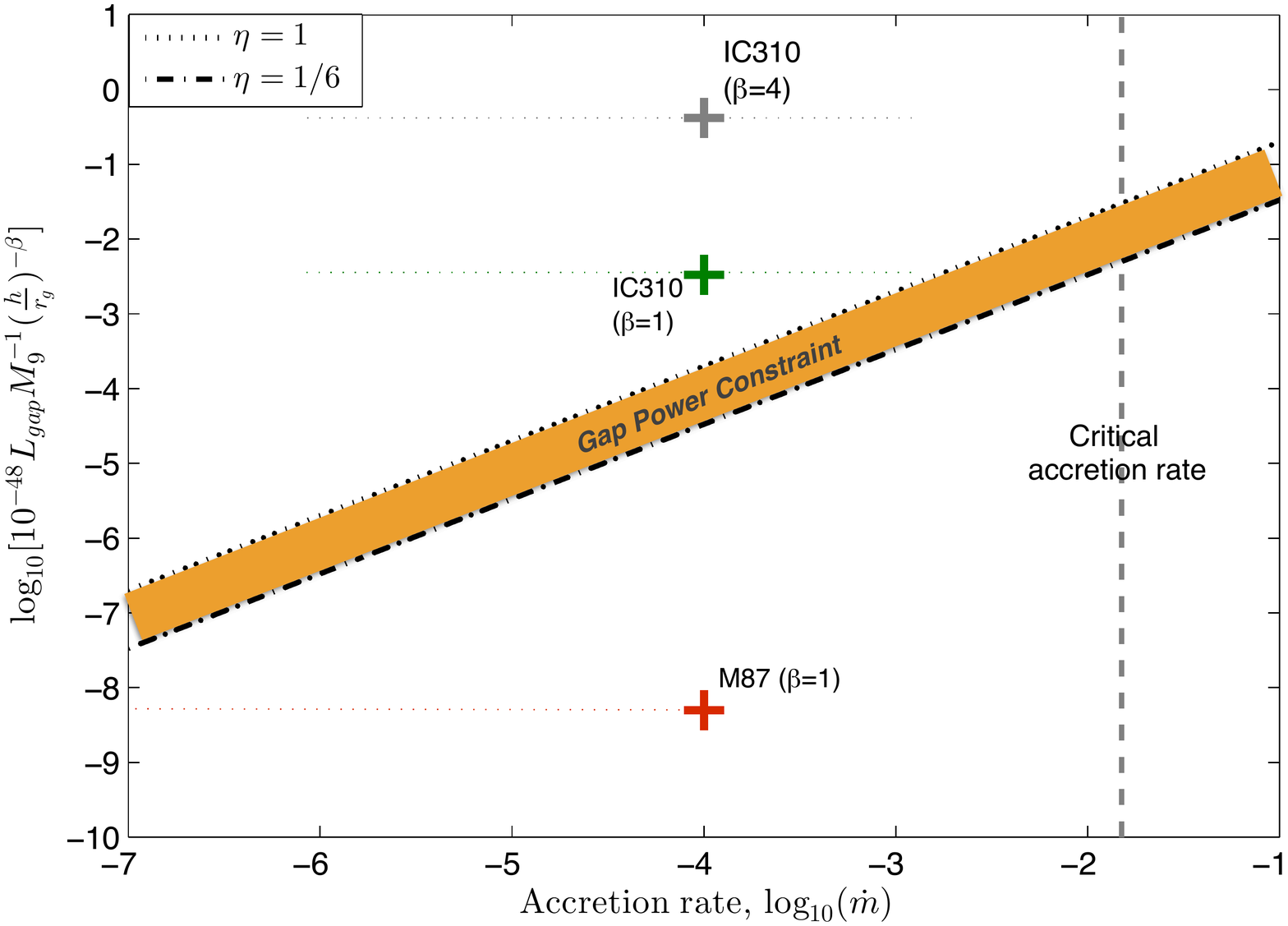}
\caption{Maximum possible gamma-ray power (orange band) of a magnetospheric gap as a function of accretion rate, cf. 
eq.~(\ref{power}). Different assumptions on the gap potential, eqs.~(\ref{potential1})-(\ref{potential2}), lead to a difference 
in extractable powers. The required transparency of the accretion environment to VHE photon introduce an upper limit 
$\dot{m}_c \simeq 0.01$ on possible rates. If the observed VHE variability is used to constrain the gap height $h$, the 
extractable power is in principle sufficient to account for the gamma-ray emission seen from M\,87, yet under-predicts 
the VHE emission seen from IC\,310. See ref.~\cite{kat17} for further details.}\label{fig2}
\label{default}
\end{center}
\end{figure}

The detection of magnetospheric $\gamma$-ray emission features in principle allows for a fundamental probe of the near black 
hole environment including accretion physics and jet formation. In reality, however, this may be more difficult to achieve. On the
one hand, to allow for the escape and detectability of VHE gamma-rays the inner accretion flow needs to be radiatively inefficient
as otherwise severe $\gamma\gamma$-absorption in the disk photon field will occur. On the other hand, for classical blazar-type 
sources (with small jet inclination) magnetospheric emission is likely to be overpowered by the strongly Doppler-boosted emission 
of their jets. Only if the jet is sufficiently misaligned such that Doppler effects are modest may we expect to see it. This has made 
misaligned and under-luminous AGN, in particular {\it nearby radio galaxies}, to the most promising targets, see e.g.,~\cite{rie17} 
for review and further references. 
The occurrence of magnetospheric emission in these sources could possibly account for the spectral hardening at GeV energies 
seen in some sources \cite{sah13,bro17} and under suitable conditions for rapid VHE $\gamma$-ray variability on horizon 
crossing-times and shorter \cite{alek,hir16}. For a moderately massive and rather weak gamma-ray source such as, e.g., {\it 
Centaurus A} (at $d \simeq 3.7$ Mpc), with $M_{BH} \simeq (0.5-1)\times 10^8 M_{\odot}$ and characteristic (isotropic equivalent) 
$L_{\rm VHE} \sim 3 \times 10^{39}$ erg/s, variability on timescales $\Delta t_H \sim r_g/c=8.3~(M_{BH}/10^8\,M_{\odot})$ min is unlikely 
to become detectable given current sensitivities, i.e., the source may appear as quasi-steady on the required instrumental integration 
times (no evidence for significant short-term variability has in fact been reported so far, neither in the HE nor the VHE domain), with 
magnetospheric emission probably only becoming apparent as an additional contribution towards higher energies. The situation can 
be different for more massive (larger $r_g$) and/or more luminous gamma-ray sources. For the radio galaxy {\it M\,87} (at $d\simeq 
16.7$ Mpc) with a black hole mass of $M_{BH} \simeq (2-6)\times 10^9 M_{\odot}$, corresponding to crossing times of $\Delta t_H 
\simeq r_g/c \simeq (3-9)$ hr, day-scale VHE variability has been detected during several TeV high states (where $L_{\rm VHE}\sim 
10^{41}$ erg/s). There are indications that the TeV emission is accompanied by delayed radio core flux enhancements supporting 
the conclusion that the VHE emission may originate at the jet base very close to the black hole, e.g., \cite{acc09}. One typically expects 
magnetospheric gaps to possess maximum heights of $h\lesssim r_g$ (if efficient pair production takes place, screening may well
occur earlier, cf.~\cite{lev11}). The (so far) observed day-scale variability in M\,87 thus does not impose severe constraints on $h$,
cf. eq.~\ref{power}. However, to ensure transparency to VHE photons the accretion flow needs to be radiatively inefficient (of ADAF-
type) and this on average constrains inner accretion rates to satisfy $\dot{m}_c \lesssim 0.01$, see e.g., \cite{kat17} for details. When
this is put in context, extractable gap powers are such as to allow accommodation of the gamma-ray emission seen from M\,87, see
Fig.~\ref{fig2}. A magnetospheric contribution could also account for the apparent VHE excess above a simple Fermi-LAT power 
law extrapolation, cf.~\cite{rie12} for details and discussion. 
The situation is different for the Perseus Cluster radio galaxy {\it IC\,310} (at $d \sim 80$ Mpc), believed to host a black hole of 
$M_{BH} \simeq 3\times 10^8 M_{\odot}$. The minute-scale VHE variability that has been seen during a strong VHE flare in 
Nov. 2012 (with isotropic $L_{\rm VHE} \sim 2\times 10^{44}$ erg/s) would imply a gap height $h \lesssim c ~\Delta t 
\simeq 0.2 r_g$ \cite{alek}. When this is employed in eq.~\ref{power} along with the ADAF constraint $\dot{m}_c \simeq 0.01$, 
extractable powers tend to becomes too small to account for the observed VHE emission, see Fig.~\ref{fig2}, thus disfavouring 
conventional magnetospheric scenarios for its origin.

\section{Conclusions}
Non-thermal magnetospheric processes could in principle lead to a non-negligible contribution at gamma-ray energies that
could vary on horizon crossing times and introduce specific spectral features (e.g., hardening in the overall source spectrum). 
As this contribution is non-boosted, potential targets would need to be close enough and possess jets sufficiently misaligned 
for this emission to become detectable by current instruments. The extractable gap power is a sensitive function of the gap 
height $h$, $L_{\rm gap} \propto L_{\rm BZ}\,(h/r_g)^{\beta}$, where $L_{\rm BZ}$ denotes the characteristic Blandford-Znajek 
jet power and $\beta \geq 1$ defines the gap potential that is realised. Detection of rapid variability on timescales smaller than 
$r_g/c$ thus becomes most constraining and could allow to probe different gap descriptions. When put in context of recent 
observations, this suggests that the gamma-ray emission seen from M\,87 may have a magnetospheric origin, while such a 
scenario appears rather disfavoured in the case of IC\,310.  The detection of magnetospheric emission generally allows for a 
sensitive probe of the near black hole environment and is thus of prime interest for advancing our understanding of the physical 
processes in extreme environments.

\end{document}